\documentclass{article}
\usepackage{spconf,amsmath,graphicx}

\usepackage{amsfonts}
\usepackage{multirow}
\usepackage{hyperref}
\usepackage{subfigure}
\usepackage{booktabs,setspace}
\usepackage{threeparttable}
\ninept
\usepackage{setspace}

\usepackage{amssymb,CJK, indentfirst,balance,appendix}

\title{M2MeT: THE ICASSP 2022 MULTI-CHANNEL MULTI-PARTY MEETING TRANSCRIPTION CHALLENGE}

\name{\begin{tabular}{c}Fan Yu$^{1,4}$, Shiliang Zhang$^{1}$,Yihui Fu$^4$, Lei Xie$^{4*}$\thanks{* Lei Xie is the corresponding author.}, Siqi Zheng$^1$, Zhihao Du$^1$, \\ 
Weilong Huang$^1$, Pengcheng Guo$^4$, Zhijie Yan$^1$, Bin Ma$^2$, Xin Xu$^3$, Hui Bu$^3$\end{tabular}}

\address{$^1$Speech Lab, Alibaba Group, China \\
        $^2$Speech Lab, Alibaba Group, Singapore \\
        $^3$Beijing Shell Shell Technology Co., Ltd., Beijing, China \\
        $^4$AISHELL Foundation
        }
\begin{document}

%
\maketitle

\begin{abstract}
Recent development of speech processing, such as speech recognition, speaker diarization, etc., has inspired numerous applications of speech technologies. The meeting scenario is one of the most valuable and, at the same time, most challenging scenarios for the deployment of speech technologies. Specifically, two typical tasks, speaker diarization and multi-speaker automatic speech recognition have attracted much attention recently. However, the lack of large public meeting data has been a major obstacle for the advancement of the field. Therefore, we make available the \emph{AliMeeting} corpus, which consists of 120 hours of recorded Mandarin meeting data, including far-field data collected by 8-channel microphone array as well as near-field data collected by headset microphone. Each meeting session is composed of 2-4 speakers with different speaker overlap ratio, recorded in rooms with different size. Along with the dataset, we launch the ICASSP 2022 Multi-channel Multi-party Meeting Transcription Challenge (M2MeT) with two tracks, namely speaker diarization and multi-speaker ASR, aiming to provide a common testbed for meeting rich transcription and promote reproducible research in this field. In this paper we provide a detailed introduction of the AliMeeting dateset, challenge rules, evaluation methods and baseline systems.

\end{abstract}

\vspace{-0.1cm}
\section{Introduction}
\label{sec:intro}
\vspace{-0.1cm}

Meeting transcription is the task to address the ``who speaks what at when'' problem in multi-speaker meetings. The task is considered as one of the most challenging tasks in speech processing due to free speaking style and complex acoustic conditions, such as overlapping speech, unknown number of speakers, far-field signals in large conference rooms, noise and reverberation etc. As a result, tackling the problem requires a well designed speech system with functionalities from multiple related speech processing components including but not limited to front-end signal processing, speaker identification, speaker diarization and automatic speech recognition, etc.

The lack of sizable labeled meeting data is one of the main bottlenecks to the current development of advanced meeting transcription system. Design of such a corpus desires both complex room and recording setup as well as time-consuming human transcription.
Since meeting transcription involves numerous related processing components, more information has to be carefully collected and labelled, such as speaker identity, speech context, onset/offset time, etc. All the information requires precise and accurate annotations, which is expensive and time-consuming.
Although several relevant datasets have been released~\cite{chen2020continuous,watanabe2020chime,janin2003icsi,mostefa2007chil,renals2008interpretation}, as discussed in ~\cite{fu2021aishell}, most of them suffer from various limitations, ranging from corpus setup, such as corpus size, number of speakers, variety of spatial locations relative to the microphone arrays, collection condition, etc., to corpus content, such as recording quality, accented speech, speaking style, etc. Moreover, almost all publicly available meeting corpora are collected in English, and the differences among languages limit the development of Mandarin meeting transcription. A publicly available large real-recorded Mandarin speech dataset in meeting scenario is in its great necessity. A common evaluation platform to promote research in this field is also necessary. 

In this paper, we make available \textit{AliMeeting}, a sizeable Mandarin meeting corpus, containing 120 hours real meeting data recorded by 8-channel directional microphone array and headset microphone.
To the best of our knowledge, AliMeeting and Aishell-4 are currently the only publicly available meeting datasets in Mandarin.
The recently released Aishell-4~\cite{fu2021aishell} covers a variety of aspects in real-world meetings, including diverse recording conditions, various number of meeting participants, various overlap ratios and noise types, and meets the modern requirement of accurate labeling.
On the basis of these advantages, AliMeeting increases the number of speakers and meeting venues to enhance generalization, while particularly adding the multi-talk discussion with a high speaker overlap ratio in the meetings. High-quality transcriptions on multiple aspects are provided for each meeting, allowing researchers to explore different aspects of meeting processing.
Furthermore, the AliMeeting corpus provides not only far-field data collected by 8-channel microphone array but also near-field data collected by the headset microphone which may benefit data augmentation.
Noticeably, our AliMeeting dataset will be presented along with Aishell-4~\cite{fu2021aishell} as M2MeT Challenge \footnote{Challenge website: https://www.alibabacloud.com/m2met-alimeeting} training data for researchers to better facilitate research on meeting transcription.

Meeting transcription usually contains at least three major subtasks, namely front-end processing, speaker diarization and speech recognition. Firstly, speech enhancement~\cite{hu2020dccrn} and separation~\cite{luo2020dual}, also known as front-end processing, aims to handle the complex acoustic environment including noisy and speech overlapped scenarios. Thanks to the rapid development of deep learning in recent years, speech enhancement and separation have received remarkable progress and consistent improvement on public datasets, e.g., WSJ0-2mix~\cite{hershey2016deep}, VoiceBank-DEMAND~\cite{valentini2016investigating} and REVERB Challenge~\cite{kinoshita2016summary}. 
However, in real meeting scenarios, there are still many problems to be solved, such as noise, reverberation and signal attenuation, overlapping mismatch between real scenarios and most hypotheses in current research, etc.
Secondly, speaker diarization is the task to label the audio with classes corresponding to speaker identity. Traditional speaker diarization usually contains an embedding extraction, and a clustering step, where the input audio stream is firstly converted into speaker-specific representation~\cite{wang2018speaker,snyder2018x}, followed by a clustering process that aggregates the regions of each speaker into separated clusters.
Thanks to the powerful modeling capabilities of deep learning, the extraction of speaker embedding using neural networks is the most widely adopted method in recent years, such as d-vectors~\cite{wang2018speaker} and i-vectors~\cite{snyder2018x}. Recently, researchers proposed end-to-end neural diarization (EEND)~\cite{fujita2019end2}, which replaces the individual sub-modules in traditional speaker diarization systems mentioned above with one neural network.
Nevertheless, the problem of generalization of mismatch between real data and synthetic data, the counting problem of variable number of speakers, and the poor performance of overlapping speech recognition are still huge challenges in real meeting scenarios.
Finally, the accuracy and efficiency of ASR have an important impact on user experience in meeting transcription. A drastic improvement in accuracy and robustness has been brought to modern single speaker ASR system~\cite{dong2018speech,gulati2020conformer}, and several works show promising results in multi-speaker ASR~\cite{kanda2020serialized,guo2021multi} by employing synthetic datasets for experimentation. However, the problem of unknown number of speakers, variable overlap rate, and accurate speaker identity assignment for utterances are still considered unsolved, especially for real-world applications. As stated in ~\cite{kanda2020serialized}, ASR for multi-speaker speech remains to be a very difficult problem especially when speech utterances from multiple speakers significantly overlap in monaural recordings.


In order to inspire research on advanced meeting rich transcription, we propose the Multi-channel Multi-party Meeting Transcription Challenge (\textit{M2MeT}) based on the new AliMeeting corpus introduced above as well as the recently-released Aishell-4 corpus~\cite{fu2021aishell}. We particularly set up two main challenge tracks of speaker diarization and multi-speaker ASR. Meanwhile, we also set up two sub-tracks for both tracks: the first sub-track limits the usage of data while the second sub-track allows the participants to use extra constrained data. Challenge rules and baseline systems are also released along with the corpus.


\begin{figure}[t]
	\centering
	\includegraphics[scale=0.28]{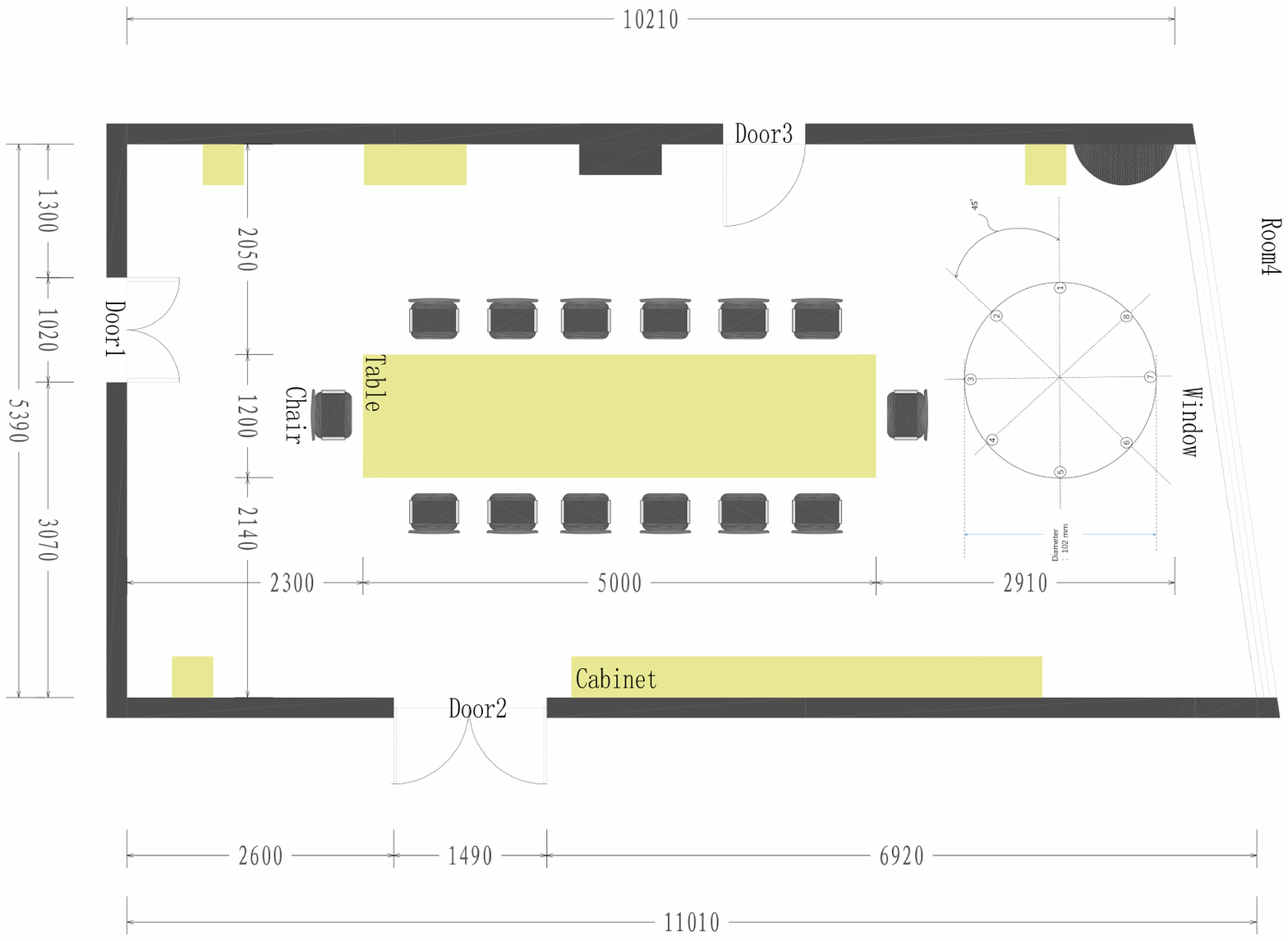}
	\vspace{-0.3cm}
	\caption{
		An example of recording venue and the topology of microphone array.
	}
	\vspace{-0.6cm}
	\label{meeting_room}
\end{figure}

\vspace{-0.3cm}
\section{The AliMeeting Corpus}
\label{sec:Datasets}
\vspace{-0.2cm}
AliMeeting contains 118.75 hours of speech data in total. The dataset is divided into 104.75 hours for training (Train), 4 hours for evaluation (Eval) and 10 hours as test set (Test) for challenging scoring and ranking. Specifically, the Train and Eval sets contain 212 and 8 sessions, respectively. Each session consists of a 15 to 30-minute discussion by a group of participants. The total number of participants in Train and Eval sets is 456 and 25, with balanced gender coverage. The Train and Eval sets will be released to the participants at the beginning of the challenge, while the Test set audio will be released at the final challenge scoring stage.

The dataset is collected in 13 meeting venues divided into three types: small, medium, and large rooms, whose sizes range from 8 to 55 m$^{2}$. Different rooms give us a variety of acoustic properties and layouts. The detailed parameters of each meeting venue will be released together with the Train data. The type of wall material of the meeting venues covers cement, glass, etc. Other furnishings in meeting venues include sofa, TV, blackboard, fan, air conditioner, plants, etc. During recording, the participants of the meeting sit around the microphone array which is placed on the table and conduct a natural conversation. The microphone-speaker distance ranges from 0.3 to 5.0 m. All participants are native Chinese speakers speaking Mandarin without strong accents. During the meeting, various kinds of indoor noise including but not limited to clicking, keyboard, door opening/closing, fan, bubble noise, etc., are made naturally. For both Train and Eval sets, the participants are required to remain in the same position during recording. There is no speaker overlap between the Train and Eval set. An example of the recording venue from the Train set is shown in Fig.~\ref{meeting_room}.

The number of participants within one meeting session ranges from 2 to 4. To ensure the coverage of different overlap ratios, we select various meeting topics during recording, including medical treatment, education, business, organization management, industrial production and other daily routine meetings. The average speech overlap ratio of Train and Eval sets are 42.27~\% and 34.76~\%, respectively. More details of AliMeeting are shown in Table~\ref{tab:info}. A detailed overlap ratio distribution of meeting sessions with different numbers of speakers in the Train and Eval set is shown in Table~\ref{tab:overlap}.

\vspace{-0.5cm}
\begin{table}[!htb]
\centering

\caption{Details of AliMeeting Tran and Eval data.}
\vspace{0.15cm}
\begin{tabular}{ccc}
\toprule
                     & Train & Eval \\ \hline
Duration (h)         & 104.75   & 4.00      \\
\#Session            & 212      & 8         \\
\#Room               & 12       & 5          \\
\#Participant        & 456      & 25         \\
\#Male               & 246      & 12         \\
\#Female             & 210      & 13         \\
Overlap Ratio (Avg.) & 42.27\%  & 34.20\%    \\ 
\bottomrule
\end{tabular}
\label{tab:info}
\vspace{-0.7cm}
\end{table}

\begin{table}[!htb]
\centering
\label{tab:overlap}
\caption{Statistics on the number of speakers (\#Speaker), speaker overlap ratio (OR\%) and the number of meeting sessions (\#Session) for Train and Eval sets.}
\vspace{0.15cm}
\begin{tabular}{ccccc}
\toprule
\multicolumn{1}{c}{\multirow{2}{*}{\#Speaker}} & \multicolumn{2}{c}{Train}    & \multicolumn{2}{c}{Eval} \\  \cline{2-5} 
\multicolumn{1}{c}{}    & OR(\%) & \#Session      & OR(\%) & \#Session \\\hline
2              & 14.65 & 71  & 15.80 & 3 \\
3              & 32.43  & 15  & 27.21  & 1 \\
4              & 59.00  & 126 & 49.75  & 4 \\\hline
2,3,4          & 42.27  & 212 & 34.20  & 8 \\
\bottomrule
\end{tabular}
\vspace{-0.1cm}

\end{table}

We also record the near-field signal of each participant using a headset microphone and ensure that only the participant's own speech is recorded and transcribed. It is worth noting that the far-field audio recorded by the microphone array and the near-field audio recorded by the headset microphone will be synchronized to a common timeline range.
All scripts of the speech data are prepared in TextGrid format for each session, which contains the information of the session duration, speaker information (number of speaker, speaker-id, gender, etc.), the total number of segments of each speaker, the timestamp and transcription of each segment, etc. We also provide unprocessed raw text in which non-speech events are transcribed as well, such as pauses, noise, overlap, etc.

\vspace{-0.4cm}
\section{TRACK SETTING AND EVALUATION}
\label{sec:track_and_evaluation}
\vspace{-0.1cm}
\subsection{Speaker Diarization (Track 1):}
\vspace{-0.1cm}
Speaker diarization, also known as speaker segmentation and clustering, addresses the “who spoke when” problem by logging speaker-specific speech events on multi-speaker audio data. 
The AliMeeting Train and Eval data sets provide not only the audio of the far-field meeting scene recorded by the microphone array with multiple speakers talking, but also the near-field audio recorded by each participant's headset microphone.
The Test set, which includes 10 hours of meeting data, will be released shortly (according to the timeline) for challenge scoring and ranking. Meanwhile, the organizers will only provide the far-field audio in the Test set recorded by 8-channel microphone array and the corresponding sentence segmentation timestamp. 
The sentence segmentation timestamp is generated by merging overlapping segments into one speech segment until no overlapping segment exist. 
Note that the organizers will not provide the headset near-field audio and the transcriptions. Participants need to determine the speaker at each time point, and an RTTM file needs to be provided for each session.

The accuracy of speaker diarization system in this track is measured by Diarization Error Rate (DER)~\cite{fiscus2006rich} where DER is calculated as: the summed time of three different errors of speaker confusion (SC), false alarm (FA) and missed detection (MD) divided by the total duration time, as shown in 
\vspace{-0.1cm}
\begin{equation}\label{eq:DER-COMPUTE}
 \text{DER} = \frac {\mathcal T_{\text{SC}} + \mathcal T_{\text{FA}} + \mathcal T_{\text{MD}}}{\mathcal T_{\text{Total}}} \times 100\%,
 \vspace{-0.2cm}
\end{equation}
where $\mathcal T_{\text{SC}}$, $\mathcal T_{\text{FA}}$ and $\mathcal T_{\text{MD}}$ are the time duration of the three errors, and $\mathcal T_{\text{Total}}$ is the total time duration.

Hungarian algorithm~\cite{kuhn1955hungarian} is adopted to establish time alignment between the hypothesis outputs and the reference transcript. In order to mitigate the effect of inconsistent annotations and human error in reference transcript, the Rich Transcription 2006 evaluation~\cite{fiscus2006rich} sets a 0.25 second ``no score" collar around every boundary of the reference segment. Since this evaluation scheme has been widely adopted in the literature~\cite{park2021review}, we follow this setup in the challenge.

\vspace{-0.3cm}
\subsection{Multi-Speaker ASR (Track 2):}
\label{Multi-Speaker}
\vspace{-0.1cm}
The challenge of multi-speaker ASR is to handle overlapped speech and to recognize the content of multiple speakers. Meanwhile, the provided audio of final test set (Test) is the same as Track 1. Finally, participants are required to transcribe each speaker, but are not required to identify a corresponding speaker for each transcript.

The accuracy of multi-speaker ASR system in Track 2 is measured by Character Error Rate (CER). The CER compares, for a given hypothesis output, the total number of characters, including spaces, to the minimum number of insertions (Ins), substitutions (Subs) and deletions (Del) of characters that are required to obtain the reference transcript. Specifically, CER is calculated by:
\vspace{-0.1cm}
\begin{equation}\label{eq:CER-COMPUTE}
\vspace{-0.1cm}
 \text{CER} = \frac {\mathcal N_{\text{Ins}} + \mathcal N_{\text{Subs}} + \mathcal N_{\text{Del}} }{\mathcal N_{\text{Total}}} \times 100\%,
\end{equation}
where $\mathcal N_{\text{Ins}}$, $\mathcal N_{\text{Subs}}$, $\mathcal N_{\text{Del}}$ are the character number of the three errors, and $\mathcal N_{\text{Total}}$ is the total number of characters. 
Considering the permutation invariant training (PIT) problem, we propose two schemes to calculate CER of the overlapping speech. 
First, we sort the reference labels according to the start time of each utterance and join the utterances with the $\left \langle \text{sc}\right \rangle$ token, which called utterance-based first-in first-out (FIFO) method. The second methods is based on speaker, where utterances from the same speaker are combined, and then we will calculate all possible concatenation patterns.

\vspace{-0.3cm}
\subsection{Sub-track Arrangement}
\vspace{-0.2cm}
For both tracks, we also set up two sub-tracks:
\vspace{-0.1cm}
\begin{itemize}
\item Sub-track I: Participants can only use constrained data to build both systems, and the usage of extra data is strictly prohibited. In other words, system building for Track 1 and Track2 are restricted to AliMeeting, Aishell-4~\cite{fu2021aishell} and CN-Celeb~\cite{fan2020cn}.
\item Sub-track II: Besides the constrained data, participants can use any data set publicly available, privately recorded, and manual simulation for system building. However, the participants have to clearly list the data used in the final system description paper. If manually simulated data is used, please describe the data simulation scheme in detail.
\end{itemize}
\vspace{-0.1cm}

\vspace{-0.3cm}
\section{Baseline systems}
\vspace{-0.1cm}
We will release baseline systems \footnote{Baseline systems: https://github.com/yufan-aslp/AliMeeting} along with the Train and Eval data for quick start and reproducible research. For the multi-channel data of AliMeeting and Aishell-4 recorded by far-field microphone array, we selected the first channel to obtain \textit{Ali-far} and \textit{Aishell-far}, and adopted CDDMA Beamformer~\cite{huang2020differential,zheng2021real} to generate \textit{Ali-far-bf}, while \textit{Ali-near} is the single-channel data recorded by headset microphone. Note that the following training and evaluation in our experiments are all based on the single-channel data (\textit{Ali-near}, \textit{Ali-far}, \textit{Ali-far-bf} and \textit{Aishell-far}). We further use prefix \textit{Train-} and \textit{Eval-} to indicate different sets, e.g., Train-Ali-far-bf denotes the one channel data outputted by the beamformer which takes the AliMeeting 8-channel array Train data as input.

\vspace{-0.3cm}
\subsection{Speaker Diarization Baseline}
\label{sec:BASELINE}
\vspace{-0.1cm}


We adopt the Kaldi-based diarization system from the CHiME-6 challenge as the baseline system~\cite{watanabe2020chime}. The diarizaiton module includes speaker embedding extractor and clustering. It is worth noting that we will provide segmentation timestamp to participants to obtain oracle SAD labels. So we don't need to provide SAD module. 


The speaker embedding network is based on ResNet~\cite{zeinali2019but} which is trained using CNCeleb~\cite{fan2020cn}.
The speaker embedding network is trained using stochastic gradient descent (SGD) and additive angular margin loss.
We ramp-up the margin during the first two epochs and then train the network for the following epoch with a fixed margin $m$ = 0.2. 
And data augmentation is performed in the same way as the SRE16 Kaldi recipe~\footnote{https://github.com/kaldi-asr/kaldi/blob/master/egs/sre16/v2}. 
The feature fed into the speaker embedding network is 64-dimensional mel-filterbank which is extracted every 15 ms with 25 ms window. 
The post processing result of speaker embedding is derived from probabilistic linear discriminant analysis (PLDA)~\cite{ioffe2006probabilistic} model which is trained on the same data as the speaker embedding network. 

All speaker embeddings generated will be pre-clustered using agglomerative hierarchical cluster (AHC)~\cite{han2008strategies} algorithm to obtain the initial speaker labels.
The threshold of AHC algorithm is set to 0.015. 
The dimension of speaker embedding is reduced from 256 to 128 for further clustering using the Variational Bayesian HMM clustering (VBx)~\cite{landini2020bayesian} model. 
Thus, the segmented utterances from the whole meeting session together with their corresponding speaker information.





\begin{table}[!htb]
\centering
\vspace{-0.5cm}
\caption{Speaker diarization results of meeting session with different number of speaker on Eval set (DER\%).}
\vspace{0.15cm}
\begin{threeparttable}[t]
\begin{tabular}{ccc}
\toprule
\#Speaker & Eval-Ali-far & Eval-Ali-far-bf \\ \hline
Collar size (s)  &0.25 / 0 & 0.25 / 0  \\ 
\toprule
2,3,4                      & 15.24 / 24.52       &  15.46 / 24.67          \\ \hline
2                          & 4.84 / 9.32       & 4.96 / 9.41          \\
3                          & 10.90 / 19.03       & 10.70 / 18.94          \\
4                          & 23.18 / 34.12       & 23.40 / 34.54          \\ 

\bottomrule

\end{tabular}

\end{threeparttable}
\vspace{-0.3cm}
\label{tab:speakers}
\end{table}

DER is scored here with collar size of 0 and 0.25 second, but the challenge evaluation only compares the results of 0.25 second collar size. 
As shown in Table~\ref{tab:speakers}, since our speaker diarization system does not take overlapped speech into consideration as we assumed that each speech frame corresponds to only one of the speakers, DER is higher in the meeting scene with more speakers and higher overlap ratio. 
Due to a large amount of interludes in the AliMeeting data, there is lots of short-time overlap around boundary of the speaker segments, so DER is particularly high at 0 second collar size.
Considering the high overlap rate of AliMeeting data, beamformer will suppress the voice of the secondary speaker, resulting in the speaker diarization performance of the overlap segments is not improved.

\vspace{-0.3cm}
\subsection{Multi-Speaker ASR Baseline}
\vspace{-0.1cm}
We also use a transformer-based end-to-end model~\cite{DBLP:journals/corr/VaswaniSPUJGKP17} as our single speaker baseline, which is based on an encoder-decoder structure with an attention mechanism. 
Meanwhile, In order to extract fine-grained local features well, our encoder block follows the same Conformer architecture as in~\cite{gulati2020conformer, guo2021recent}, which includes a multi-head self-attention (MHSA) module, a convolution (Conv) module, and a pair of position-wise feed-forward (FFN) module in the Macaron-Net style.
In addition, incorporating the relatively positional embeddings to MHSA modules further improves the generalization of the model on variable lengths of input sequences. 

To obtain better recognition of multi-party meetings, we adopt the Serialized Output Training (SOT) method in~\cite{kanda2020serialized}, which generates the transcriptions of multiple speakers one after another. The SOT method has an excellent ability to model the dependencies among outputs for different speakers and no longer has a limitation on the maximum number of speakers. Remarkably, we still employ Transformer model with conformer encoder. To recognize multiple utterances, SOT serializes multiple references into a single token sequence. For example, for a two-speaker case, the reference label will be given as $R=\{r_1^1,..,r_{N_1}^1,\left \langle \text{sc}\right \rangle,r_1^2,..,r_{N_2}^2,\left \langle \text{eos}\right \rangle\}$, note that $\left \langle \text{eos}\right \rangle$ is used only at the end of the entire sequence and the special symbol $\left \langle \text{sc}\right \rangle$ represent the speaker change which is used to concatenate different utterances. 
In order to avoid the complex calculation of PIT for all possible concatenation patterns, we adopt utterance-based FIFO method which described in Section~\ref{Multi-Speaker}.
As the experiment shows in \cite{kanda2020serialized} that the FIFO method achieved a better CER than the method of calculating all permutations. Based on this result, we used the FIFO method for our all ASR experiments.

In all experiments, we use 71-dimensional mel-filterbanks feature as the input of the acoustics model and frame length is 25 ms with a 10 ms shift. We use the ESPnet~\cite{Watanabe2018ESPnet} end-to-end speech processing toolkit to build our ASR system. We follow the standard configuration of the state-of-the-art ESPnet Conformer structures which contains a 12-layer encoder and 6-layer decoder, where the dimension of attention and feed-forward layer is set to 256 and 2048 respectively. In all attention sub-layers, the head number multi-head attention is set to 4. The whole network is trained for 100 epochs and warmup is used for the first 25,000 iterations. We use 4950 commonly used Chinese characters as the modeling units.

\begin{table}[]
\setlength\tabcolsep{1.5pt}
\centering
\vspace{-0.2cm}
\caption{Single speaker ASR results comparison on Eval set (CER\%).}
\vspace{0.15cm}
\footnotesize
\begin{threeparttable}[t]
\begin{tabular}{clccc}
\toprule
Conformer & \multicolumn{1}{c}{Training data}     & Eval-Ali-near & Eval-Ali-far & Eval-Aishell-far \\ \hline
A         & Train-Ali-near         & 9.9     & 49.0    & 32.9        \\
B         &  \;+Train-Aishell-far     & 9.3     & 37.5    & 28.1        \\ \hline
C         & \; \;+External data & \textbf{7.8}     & 35.0    & 24.7        \\
D         & \; \;+External data$^*$ & 8.2     & \textbf{33.3}    & \textbf{23.7}       \\
\bottomrule
\end{tabular}
\begin{tablenotes}
    \footnotesize
        \item  External data includes Aishell-1~\cite{aishell_2017}, AIDataTang\footnote{http://www.openslr.org/62/} and ST-CMD\footnote{http://www.openslr.org/38/}.
		\item  {$^*$}: Noise and reverberation simulated.
\end{tablenotes}
\end{threeparttable}
\label{tab:asr_single}
\vspace{-0.6cm}
\end{table}

The results of single speaker ASR are shown in Table~\ref{tab:asr_single}, where the Train data used by the various Conformer models are different. ConformerA is trained with Train-Ali-near set. In order to enhance the performance of the model for the far-field Eval set, the training of ConformerB further includes the Aishell-far training data. Note that A and B are baselines for Sub-track I. In addition, for ConformerC, we add additional Aishell-1~\cite{aishell_2017}, aidatatang\footnote{http://www.openslr.org/62/} and ST-CMD\footnote{http://www.openslr.org/38/} near-field data to confirm the improvement effect of out-domain data. Considering that the added near-field data in ConformerC and the real far-field meeting data have a great distinction in the room acoustics, ConformerD simulates noise and reverberation for the added near-field data. Here C and D belong to Sub-track II where the use of external data is not limited. Table~\ref{tab:asr_single} shows that ConformerD has an obvious improvement on the far-field set, and at the same time, the performance degradation of the near-field Eval set is very small.


\begin{table}[!htb]
\centering
\vspace{-0.6cm}
\caption{Multi-speaker ASR results on Eval set (CER\%).}
\vspace{0.15cm}
\begin{tabular}{llcc}
\toprule
Conformer & \multicolumn{1}{c}{Training data} & Eval-Ali-far & Eval-Ali-far-bf \\ \hline
A         & Train-Ali-near                    & 49.0    & 45.6       \\
SOTA      & Train-Ali-far                     & 34.3    & 38.2       \\
SOTB      &  \;+Train-Ali-near                   & \textbf{30.8}    & 33.2       \\
SOTA\_bf  & Train-Ali-far-bf                  & 41.2    & 33.6       \\
SOTB\_bf  &  \;+Train-Ali-near                   & 34.3    & \textbf{29.7}      \\
\bottomrule
\end{tabular}
\label{tab:asr_multi}
\vspace{-0.3cm}
\end{table}

The results of multi-speaker ASR are shown in Table~\ref{tab:asr_multi}, where Conformer\_SOTA is trained on the Train-Ali-far set and Conformer\_SOTA\_bf is trained on the Ali-far-bf data.
In order to enhance the acoustic modeling ability of the model, we added the training data of Train-Ali-near into model B, and it can be seen that there is a significant improvement in both Eval sets.
Compared with the single-speaker model, the SOT multi-speaker model has a significant improvement on our multi-speaker Eval set, and SOT with Beamformer (bf) also has positive benefits on the Ali-far-bf Eval set. 

\vspace{-0.4cm}
\section{Conclusions}
\vspace{-0.2cm}
In this paper, we introduce the necessity of ICASSP2022 Multi-channel Multi-party Meeting Transcription Challenge (M2MeT) and describe the datasets, tracks, evaluation, metrics and baselines for the challenge. With the released sizable open data and common testbed, we intend to advance meeting transcription and related speech processing tasks through this challenge. 




\begin{spacing}{0.1}
\bibliographystyle{IEEEbib}
\bibliography{strings,refs}
\end{spacing}

\end{document}